\documentclass[usenatbib]{mn2e}
\usepackage{amssymb,amsmath,graphicx,natbib,setspace,pifont}
\usepackage{grffile}\usepackage{epsf}
\usepackage{float}

\newcommand{\HI}{\hbox{{\sc H}{\sc i}} }

\newcommand{\NHI}{{N_{\rm HI}}}

\newcommand{\cmsq}{\,{\rm cm^{-2}}}
\newcommand{\cms}{\,{\rm cm^{-2}}}

\newcommand{\Gadget}{{\small GADGET-3} }

\newcommand{\TRAPHIC}{{\small TRAPHIC}~}
\newcommand{\Zsolar}{{\rm Z}_{\odot}}

\newcommand{\apjl}{{ApJ}}

\newcommand{\apj}{{ApJ}}

\newcommand{\mnras}{{MNRAS}}

\begin{document}

\title[The metallicity distribution of HI systems]{The metallicity distribution of HI systems in the EAGLE cosmological simulations}

\author[A.~Rahmati, B. D. Oppenheimer]{
  \parbox[t]{\textwidth}{\vspace{-1cm}
    Alireza Rahmati$^{1}$, Benjamin D. Oppenheimer$^{2}$\thanks{benjamin.oppenheimer@colorado.edu}}\\
$^1$Institute for Computational Science, University of Z\"urich, Winterthurerstrasse 190, CH-8057 Z\"urich, Switzerland\\	
$^2$CASA, Department of Astrophysical and Planetary Sciences, University of Colorado, 389 UCB, Boulder, CO 80309, USA\\
}
\maketitle

\begin{abstract} 

  The metallicity of strong $\HI$ systems, spanning from damped
  Lyman-$\alpha$ absorbers (DLAs) to Lyman-limit systems (LLSs) is
  explored between $z=5\rightarrow 0$ using the EAGLE high-resolution
  cosmological hydrodynamic simulation of galaxy formation.  The
  metallicities of LLSs and DLAs steadily increase with time in
  agreement with observations.  DLAs are more metal rich than LLSs,
  although the metallicities in the LLS column density range
  ($N_{\HI}\approx 10^{17}-10^{20} \cms$) are relatively flat,
  evolving from a median $\HI$-weighted metallicity of ${\rm Z}\la
  10^{-2} \Zsolar$ at $z=3$ to $\approx 10^{-0.5} \Zsolar$ by $z=0$.
  The metal content of $\HI$ systems tracks the increasing stellar
  content of the Universe, holding $\approx 5\%$ of the integrated
  total metals released from stars at $z=0$.  We also consider partial
  LLS (pLLS, $N_{\HI}\approx 10^{16}-10^{17} \cms$) metallicities, and
  find good agreement with \citet{wot16} for the fraction of systems
  above (37\%) and below (63\%) $0.1\Zsolar$.  We also find a large
  dispersion of pLLS metallicities, although we do not reproduce the
  observed metallicity bimodality and instead we make the prediction
  that a larger sample will yield more pLLSs around $0.1 \Zsolar$.  We
  under-predict the median metallicity of strong LLSs, and predict a
  population of ${\rm Z}<10^{-3} \Zsolar$ DLAs at $z>3$ that are not
  observed, which may indicate more widespread early enrichment in the
  real Universe compared to EAGLE.

\end{abstract}


\section{introduction}
\label{sec:intro}

Neutral hydrogen in the Universe is associated with the fundamental
processes of galaxy formation, from the accretion of gas onto
galaxies, the feeding of star formation within galaxies via neutral
atomic reservoirs, and superwind feedback enriching the circumgalactic
medium (CGM) and intergalactic medium (IGM).  Observations show that
the number density of $\HI$ absorbers and their column densities
increase closer to galaxies \citep[e.g.][]{ade03, che09, pro17}
indicating an intimate link between atomic gas and galaxy growth.
Observing techniques rely on quasar absorption line spectroscopy
toward a UV-bright background source to measure the column density of
$\HI$ absorbers \citep[e.g.][]{wey98,leh07,dan16}.  $\HI$ absorbers
with columns having significant neutral fractions, which correspond
roughly to Lyman Limit Systems (LLSs; $N_{HI}\ga 10^{17.2} \cms$),
generally have heavy element absorption indicating that this hydrogen
is enriched with the products of stellar nucleosynthesis.  The
metallicity statistics and evolution of $\HI$ provide crucial
constraints on theoretical models of the flows of gas in and out of
galaxies \citep[e.g.][]{ocv08,van12} and the census of cosmic
metals produced by stars \citep[e.g.][]{fuk04,bou07,pee14}.

The trend of metallicity with $\HI$ column density has been the
subject of a number of observational surveys that focus on specific
$\HI$ column density ranges.  In general, Lyman Limit Systems refer to
absorbers with $10^{17.2} \leq N_{\HI} < 10^{20.3} \cms$, but these
are sub-divided into normal LLSs with
$10^{17.2} \leq N_{\HI} < 10^{19.0}\ \cms$ and super LLSs (SLLSs) with
$10^{19} \leq N_{\HI} < 10^{20.3} \cms$, where damping wings begin to
appear\footnote{SLLSs are sometimes referred to as sub-DLAs}.  Damped
Lyman-$\alpha$ absorbers (DLAs), defined by their damping wings, have
column densities in excess of $10^{20.3} \cms$.  Finally, partial LLSs
(pLLSs) cover column densities $10^{16.1}-10^{17.2} \cms$.
Metallicities are significantly sub-solar and in general increase with
$\HI$ column density \citep{leh13,leh16,wot16}, although this increase
can be relatively mild or even flat as observed by \citet{fum16} for
$z\approx 2.5-3.5$ LLSs and SLLSs with typical
${\rm Z}\approx 10^{-2}\Zsolar$.

The evolution of $\HI$ metallicities across the history of the
Universe is the subject of \citet{raf14}, who determines that DLA
metallicities rise for ${\rm Z} \la 10^{-1.5}\Zsolar$ at $z\ga 4$ to
$\ga 0.1 \Zsolar$ at $z\la 1$.  SLLSs show a clear increase in from
${\rm Z} \la 10^{-2}\Zsolar$ at $z=3-4$ to $10^{-0.5}\Zsolar $ below
$z=1$ \citep{fum16}.  LLSs and pLLSs also show an increase from
${\rm Z} \approx 10^{-2}\Zsolar$ at $z\approx 2.3-3.3$ \citep{leh16}
to having median metallicities of $\approx 0.1 \Zsolar$ at $z<1$
\citep{wot16}.  This latter observation and the earlier study
\citep{leh13} show intriguing metallicity bimodality for pLLSs with
two distinct peaks at ${\rm Z}=10^{-1.9}$ and $10^{-0.3}\Zsolar$.
Increasing metallicity with time is the expectation of stellar
nucleosynthesis enriching gas that initially holds the primordial
abundances of Big Bang nucleosynthesis.

Simulation show DLAs are more often associated with gaseous reservoirs
within galaxies \citep[e.g.][]{fum11,van12b,bir14}, and their
evolution is likely related to the enrichment of the atomic
interstellar medium (ISM) by the release of heavy elements at stellar
death.  Other previous simulations that have predicted DLA
metallicities include \citet{dav07}, \citet{pon08}, \citet{tes09}, and
\citet{cen12}.  LLSs, on the other hand, have been shown to trace
lower density gas associated with the CGM outside of galaxies
\citep[e.g.][]{kat96,fau11,Rahmati14} and likely indicate energetic
stellar and supermassive black hole feedback driving materials out of
galaxies.  The large spread of LLS metallicities suggests the
tantalizing possibility that metal content is linked to physical
processes regulating galaxy growth.  For example, the pLLS metallicity
bimodality could indicate the accretion of low-${\rm Z}$ gas and the
violent ejection of high-${\rm Z}$ gas.  The FIRE cosmological
hydrodynamic zoom simulations do find this general trend in
\citet{haf17}, although clear bimodality is not reproduced and instead
a large spread of intermediate metallicities do not clearly
distinguish inflows from outflows.

This work presents $\HI$ metallicities across column density from
pLLSs through DLAs and redshift from $z=5\rightarrow 0$ using the
EAGLE (Evolution and Assembly of GaLaxies and their Environments)
cosmological hydrodynamical simulation project
\citep{sch15,cra15,mca16}.  These simulations have been shown to
successfully reproduce a variety of galaxy observables
\citep[e.g.][]{sch15, fur15, tra15, seg16a, cra17}.  Although, EAGLE
was not calibrated based on observations of the IGM/CGM, the
simulations show broad agreement with absorption line statistics
probing $\HI$ \citep{Rahmati15} and good agreement with metal ions in
the IGM \citep{Rahmati16} and around $z\approx 2$ star-forming
galaxies \citep{tur17}.  However, some metal ion statistics are
under-predicted by EAGLE, including the $z\approx 3.5$ IGM
\citep{tur16} and the higher column density metal absorbers in the IGM
\citep{Rahmati16} when using the main $100$ Mpc ``Reference''
simulation.  We use the \emph{Recal-L025N0752} high-resolution (HiRes)
volume for our exploration here, which has $8\times$ higher resolution
than the Reference EAGLE simulation, although $64\times$ less volume.
This resolution was also used in the exploration of metal absorption
in the $z\la 0.3$ CGM by \citet{opp16, opp18} using zooms including
non-equilibrium effects.  The HiRes suite of simulations, including
these zooms that were shown to produce results consistent with the
HiRes volume, indicates better agreement for observed high column
density metal absorbers than the main EAGLE simulation.  We therefore
further explore this volume for the metallicity distribution of strong
$\HI$ absorbers across the cosmic history of our Universe in this
paper.

The structure of this paper is as follows.  We introduce the EAGLE
HiRes simulation in \S\ref{sec:method} and discuss our method to
calculate $\HI$ metallicities.  We present main results in
\S\ref{sec:results} focusing first on the metallicity distribution of
$\HI$ systems \S\ref{sec:ZHIdist} and then modelling the pLLS and LLS
metallicity distributions observed by \citet{wot16} in
\S\ref{sec:wotta}.  We summarize the paper in $\S$\ref{sec:dend}.
Solar metallicity is set to $\Zsolar=0.0134$ \citep{asp09}.  

\section{Methodology} \label{sec:method}

\subsection{Simulations} \label{sec:sims}
We use the \emph{Recal-L025N0752} EAGLE HiRes cosmological
hydrodynamical simulation of galaxy formation, which uses a heavily
modified version of \Gadget~last described in \citet{spr05}.  We invite
the interested reader to find a detailed description of the code in
\citet{sch15} and \citet{cra15}.  Briefly, subgrid prescriptions for
radiative cooling \citep{wie09a}, star formation \citep{sch08},
stellar evolution and chemical enrichment \citep{wie09b}, as well as
superwind feedback associated with star formation \citep{dal12} and
black hole (BH) growth \citep{ros15} are included.  The metallicity
dependent density threshold for star formation calculated by
\citet{sch04} is used, which can alter DLA metallicities by affecting
the transition between warm, atomic and cold, molecular ISM.  EAGLE
uses the ``Anarchy'' formulation of smooth particle hydrodynamics
(SPH) \citep{schal15}, which was shown to reduce the amount of
$\HI$-traced clumps in a hot medium relative to standard SPH.
\citet{pla14} cosmological parameters are assumed.

\subsection{\HI and metallicity calculation} \label{sec:ZHIcalc} 
For calculating the simulated $\HI$ column densities we account for
the main ionizing processes that shape the distribution of neutral
hydrogen. Those include collisional ionization, which is dominant at
high temperatures, and photoionization by the metagalactic ultraviolet
background (UVB) radiation, which contributes to the bulk of hydrogen
ionization on cosmic scales, particularly at $z \gtrsim 1$
\citep[e.g.,][]{Rahmati13a}. On smaller scales and close to sources,
local radiation could be the dominant source of
photoionization. However, including the impact of local radiation even
with detailed radiative transfer simulations is hindered by other
uncertainties such as the ISM physics on small scales
(\citealp{Rahmati13b}), and is beyond the scope of this study.

We use the UVB model of \citet{HM01} to account for the mean ionizing
radiation field from quasars and galaxies. The same UVB model was also
used for calculating radiative heating/cooling rates in the
hydrodynamical simulations. Moreover, this UVB model has been shown to
reproduce results consistent with the observed column density
distribution function of $\HI$ and highly ionized ions
\citep{Rahmati13a,Rahmati15,Rahmati16}.

To account for the self-shielding of gas at $\NHI \gtrsim
10^{17}\cmsq$, we use the fitting function presented in
\citet{Rahmati13a} for calculating the photoionization rate and hence
the ionization state of hydrogen atoms. This fitting function
accurately reproduces the result from radiative transfer simulations
of the UVB and recombination radiation in cosmological density fields
using \TRAPHIC \citep{Pawlik08,Pawlik11,Raicevic13}. Moreover, because
the temperature of star-forming gas in our simulations is defined by a
polytropic equation of state that is used to limit the Jeans mass, and
therefore is not physical, for calculating collisional ionization and
photoionization rates we set the temperature of the ISM particles to
$T_{\rm{ISM}} = 10^4~\rm{K}$, which is the typical temperature of the
warm-neutral ISM.

Furthermore, to account for the conversion of atomic hydrogen to
molecular hydrogen, H$_2$, we us the empirical pressure law based on
the \citet{bli06} relation. However, we note that this procedure only
affects HI column densities higher than $10^{21.5} \cms$
\citep{Rahmati13a} and therefore not very relevant for our results in
this work.

To calculate $\HI$ column densities we use SPH interpolation and
project the $\HI$ content of the full simulation onto a 2-D grid. We
use the same projection technique to calculate the $\HI$-weighted
metallicity in each cell, using SPH smoothed metallicities of SPH
particles along the projection direction. We found that using a grid
with $10,000^2 = 10^8$ pixels for projecting the full box of the
\emph{Recal-L025N0752} simulation results in converged metallicities
and $\HI$ column densities for $\NHI \lesssim 10^{22} \cmsq$, which is
the range of $\HI$ column densities we study in this work. We use 8
slices with equal widths for calculating the $\HI$ column densities in
the full simulation box. This enables us to calculate $\HI$ column
densities as low as $\NHI \approx 10^{14} \cmsq$ without being
affected by projection effects.  \citet{Rahmati13a} checked that the
projection technique does not commonly lead to multiple overlapping
$\HI$ absorbers being combined into a single, stronger absorber, and
using multiple slices further prevents overlap affecting statistics.

\section{Results} \label{sec:results}
\subsection{Metallicity distribution of $\HI$ systems} \label{sec:ZHIdist}

Using the projection technique described in \S\ref{sec:ZHIcalc}, we
produce $\approx 10^9$ pairs of $\HI$ column densities and their
associated metallicities for each snapshot of the \emph{Recal-L025N0752}
simulation. We use this data to explore the metallicity distribution
of $\HI$ systems and its evolution.

\begin{figure*}
\centerline{\hbox{\includegraphics[width=0.48\textwidth]
             {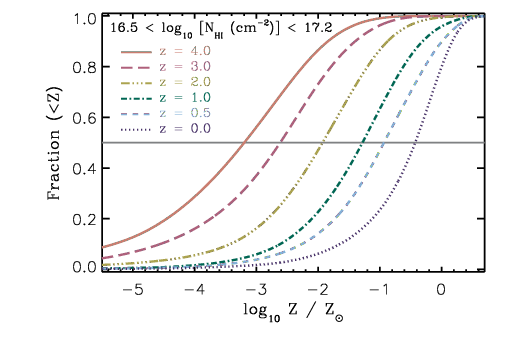}}
             \hbox{\includegraphics[width=0.48\textwidth]
             {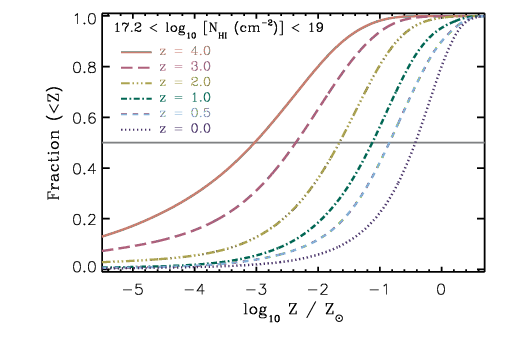}}}
\centerline{\hbox{\includegraphics[width=0.48\textwidth]
             {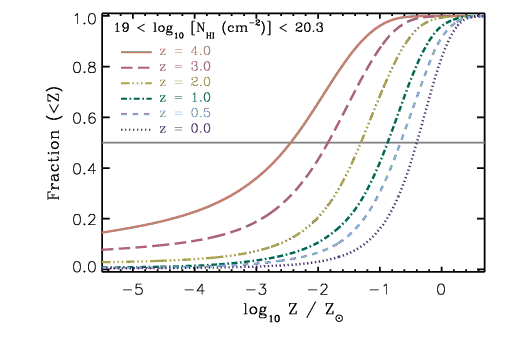}}
             \hbox{\includegraphics[width=0.48\textwidth]
             {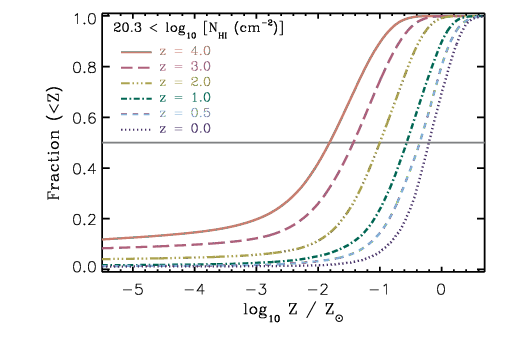}}}
\caption{The cumulative metallicity distribution functions (CMDFs) of
  $\HI$ systems at different redshifts. Panels from top-left to
  bottom-right correspond to pLLSs, LLSs, SLLSs and DLAs. The
  evolution of the CMDFs of each group of $\HI$ systems are shown
  using curves with different line-styles in each panel. The CMDFs of
  $\HI$ systems evolves strongly and the typical metallicity of all
  systems increases with decreasing redshift. The metallicity
  evolution is stronger for systems with lower $\HI$ column densities,
  which also have lower typical metallicities.}
\label{fig:ZHI-evol}
\end{figure*}

The cumulative metallicity distribution functions (CMDFs) of $\HI$
systems at different redshifts are shown in Figure
\ref{fig:ZHI-evol}. Panels in this figure from top-left to
bottom-right correspond to absorbers with different $\HI$ column
densities: pLLSs with $10^{16.5} < \NHI/\cmsq < 10^{17.2}$, LLSs with
$10^{17.2} < \NHI/\cmsq < 10^{19}$, SLLSs with $10^{19} < \NHI/\cmsq <
10^{20.3}$ and DLAs with $\NHI/\cmsq > 10^{20.3}$. The evolution of
the CMDFs of each group of $\HI$ systems are shown using curves with
different line-styles in each panel. As the panels illustrate, the
CMDFs of all considered $\HI$ systems evolves strongly and the typical
metallicity of all systems increases with decreasing redshift. Another
conclusion one can draw from this figure is that the metallicity
evolution is stronger for systems with lower $\HI$ column densities,
which also have lower typical metallicities. The CMDFs are not
symmetric in logarithmic space and have long tails towards very low
metallicities for all $\HI$ column densities, which become greater
with increasing redshift for all $\HI$ column densities.

\begin{figure}
\centerline{\hbox{\includegraphics[width=0.48\textwidth]
             {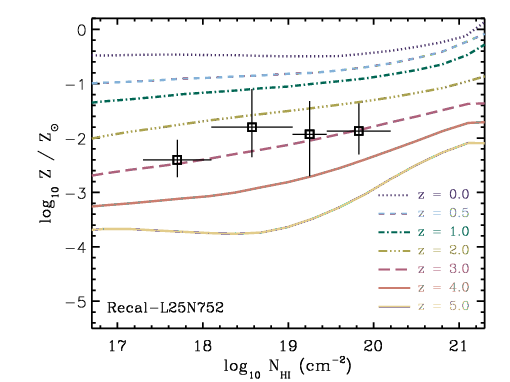}}}
\caption{Median metallicity of $\HI$ systems as a function of their
  $\HI$ column densities at different redshifts. Our predictions are
  in good agreement with observational data at $z \approx 3$ shown
  using squares with error bars taken from \citet[][their
    Fig. 15]{fum16} for a redshift range $2.5 < z < 3.5$. The
  metallicity of strong $\HI$ systems depends weakly on their $\HI$
  column density, but evolves strongly at fixed column density.}
\label{fig:ZNHI-evol}
\end{figure}

To compare our predictions against existing observational constraints
in Figure \ref{fig:ZNHI-evol}, we show the median metallicity of $\HI$
systems as a function of their $\HI$ column densities at different
redshifts. The result shows that the median metallicity of $\HI$
systems increases mildly with $\NHI$ at all redshifts but evolves much
more strongly with redshift at fixed column density. Moreover, our
predictions are in good agreement with observational data at $z
\approx 3$ shown using squares with error bars taken from
\citet[][their Fig. 15]{fum16} for a redshift range $2.5 < z < 3.5$.

\begin{figure*}
\centerline{\hbox{\includegraphics[width=0.48\textwidth]
             {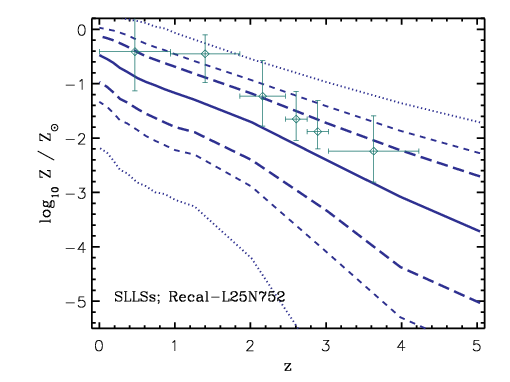}}
             \hbox{\includegraphics[width=0.48\textwidth]
             {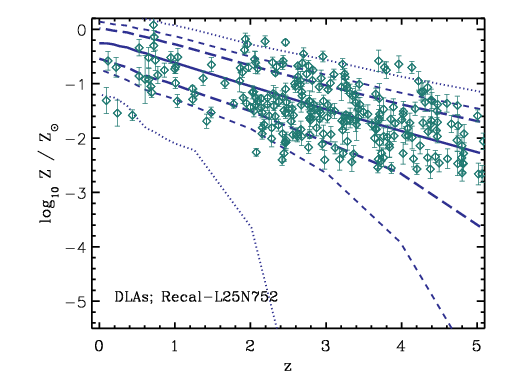}}}
\caption{Metallicities of super Lyman Limit systems (SLLSs; left
  panel) and DLAs (right panel) as a function of redshift. The solid
  curve shows the median metallicity of absorbers while long-dashed,
  dashed and dotted curves show the 25th-75th, 15th-85th and 5th-95th
  percentiles. The data points on the left panels show the
  observational constraints on the metallicity of SLLSs from
  \citet{fum16}. The data points on the right panel are a compilation
  of available observational constraints from \citet{raf14}. The
  simulation results are in reasonable agreement with observations in
  particular for DLAs while the observed metallicity of SLLSs is
  slightly underproduced in the simulations.}
\label{fig:Z-evol-DLAsLLSs}
\end{figure*}

The comparison between the metallicity distributions of DLAs and SLLSs
at different redshifts is shown in Figure
\ref{fig:Z-evol-DLAsLLSs}. The left and right panels respectively show
metallicities of SLLSs and DLAs as a function of redshifts. In each
panel, the solid curve shows the median metallicity of absorbers as a
function of redshift while long-dashed, dashed and dotted curves show
the 25th-75th, 15th-85th and 5th-95th percentiles. The data points on
the left panels show the observational constraints on the metallicity
of SLLSs from \citet{fum16}. The data points on the right panel are a
compilation of available observational constraints from
\citet{raf14}. The simulation results show overlap with observations,
particularly for DLAs at $z>2$ while the observed metallicity of SLLSs
are slightly underproduced in the simulations.  Interestingly, it
appears from the right panel of Fig. \ref{fig:Z-evol-DLAsLLSs} that
the very low-metallicity DLAs in our predictions are missing from the
observed sample, particularly at $z > 3$.

\begin{figure}
\centerline{\hbox{\includegraphics[width=0.48\textwidth]
    {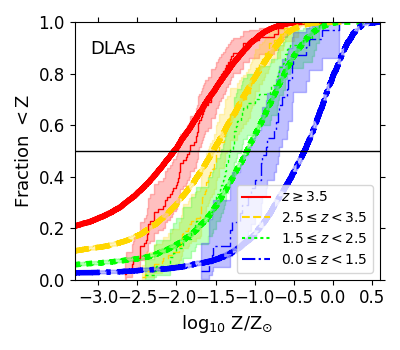}}}
\caption{CMDFs of DLAs observed by \citet{raf14} (this lines with
  shading indicating 95\% confidence limits) are split into 4 redshift
  bins.  $100$ EAGLE SMOHALOS realizations matching column density and
  redshift divided into the redshift bins are shown using thick lines
  of the matching line type to the corresponding observed line.}
\label{fig:Rafelski-SMOHALOS}
\end{figure}

To better compare the simulation to observation for DLAs, we use the
python module Simulation Mocker Of Hubble Absorption-Line
Observational Surveys (SMOHALOS) described in \citet{opp16} to
simulate the CMDFs of the \citet{raf14} sample by matching the column
density and redshift to the observed sample in Figure
\ref{fig:Rafelski-SMOHALOS}.  Random selection of column density
(within $0.05$ dex) and redshift (using the 7 redshifts plotted in
Fig. \ref{fig:ZNHI-evol}) is performed by selecting absorbers from the
entire \emph{Recal-L025N0752} volume.  We plot CMDFs of the
observations split into 4 redshift bins with thin lines and shading
corresponding to 95\% confidence limits.  The simulated CMDF
predictions are shown in thick lines without confidence limits, and
better illuminate the deviations from observed relations.  The
simulations predict more low-metallicity DLAs than observed at $z \ga
3$, but they also predict more high-metallicity DLAs than observed,
especially at super-solar levels.  The $z<1.5$ bin is the least
populated with $31$ observations, and it will be interesting to see if
further observations show the slower evolution in DLA metallicities
than our model predicts.  

\begin{figure}
\centerline{\hbox{\includegraphics[width=0.48\textwidth]
             {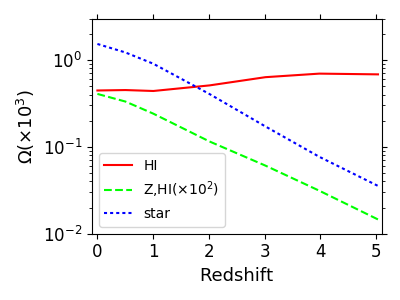}}}
\caption{The cosmic mass densities of $\HI$ (solid), $\HI$-traced
  metals (dashed, multiplied by $100$), and stars (dotted)
  as a function of redshift from $z=5\rightarrow 0$.  Total $\HI$
  remains relatively steady compared to the increase in metals, which
  follow the increase in stellar density.}
\label{fig:omega}
\end{figure}

We also sum the cosmic density of $\HI$, integrated across the $\HI$
column density distribution in Figure \ref{fig:omega} (solid line) and
compare it to the cosmic density of metals in $\HI$ (dashed line,
multiplied by $100\times$).  Both these sums are dominated by DLAs.
The increase in $\Omega_{{\rm Z},\HI}$ contrasts to the relative
invariance in $\Omega_{\HI}$, and better reflects the evolving cosmic
density of stars (dotted line) in the \emph{Recal-L025N0752}
volume.  3.2\% of all baryons are in stars by $z=0$, which compares to
0.9\% for cosmic $\HI$ mass.  The $\HI$ metallicity is $0.7 \Zsolar$
at $z=0$, and the total cosmic metal budget traced in $\HI$ gas is
$\approx 5\%$.  Far more metals remain trapped in stars and injected
into other phases of the CGM and IGM.

\subsection{Simulations of the Wotta et al. (2016) absorbers} \label{sec:wotta}

\begin{figure*}
\centerline{\hbox{\includegraphics[width=0.95\textwidth]
{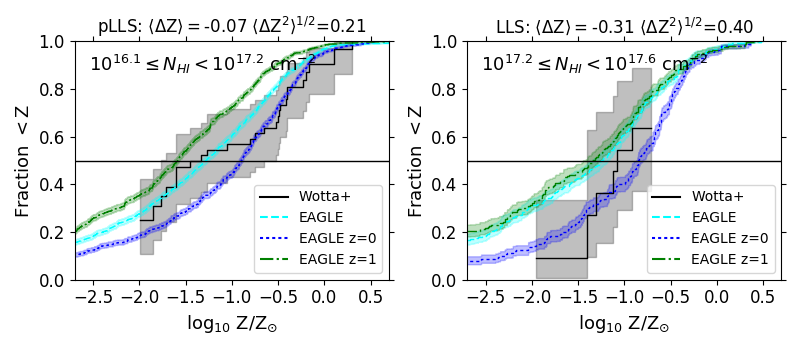}} }
\caption{CMDFs of pLLSs (left panel) and LLSs (right panel) observed
  between $z=1\rightarrow 0$ \citep[][solid]{wot16} with shading
  indicating 95\% confidence limits.  $100$ EAGLE SMOHALOS
  realizations matching column density and redshift are shown (dashed
  lines), with the mean and rms difference between simulations and
  observations in logarithmic dex listed along the top.  Realizations
  assuming all absorbers are at $z=0$ (dotted lines) and $z=1$
  (dash-dotted lines) demonstrate the predicted importance of redshift
  evolution in the simulations that is not reproduced in observations.}
\label{fig:Wotta-SMOHALOS}
\end{figure*}

We use SMOHALOS to simulate the CMDFs of observed pLLSs and LLSs of
\citet{wot16} by matching the column density and redshift of simulated
absorbers.  The CMDFs of 100 SMOHALOS realizations are shown in the
left panels of Figure \ref{fig:Wotta-SMOHALOS} for their $44$ pLLSs
(left) and $11$ LLSs (right).  Random selection of column density
(within $0.05$ dex) and redshift (within $0.06$ using 10 snapshots
between $z=1\rightarrow 0$) is performed by selecting absorbers from
the entire \emph{Recal-L025N0752} volume.  Metallicities are
calculated using $\HI$-weighted metallicities in mock spectra
generated using SpecWizard, described in \citet{sch03}, and produce
statistically consistent results with the pixel method described in
\S\ref{sec:ZHIcalc} and used throughout \S\ref{sec:ZHIdist}.  CMDFs
have shading corresponding to the $95\%$ confidence limits, which are
small for the 100 SMOHALOS realizations in these panels, but larger
for the \citet{wot16} observed distributions.  The two-sided
Kaplan-Meier estimator described in \citet{opp18} is applied to
observations to account for both upper limits due to non-detections of
metal lines and lower limits due to saturated metal lines.  These
censored measurements hardly affect the CMDFs other than truncating
the pLLSs before they reach zero at low metallicity due to upper
limits and the LLSs before they reach one at high metallicity due to
lower limits.

The simulations do not reproduce the observed pLLS bimodality, which
appears in the \citet{wot16} CMDF as a flattening around
$0.1 \Zsolar$, where very few absorbers exist.  We plot the CMDF with
confidence limits as opposed to the histogram of metallicities to show
the $2$-$\sigma$ dispersion expected in observations.  The simulated
pLLS CMDF has a median metallicity of ${\rm Z}=10^{-1.3} \Zsolar$.
The fraction of our pLLS sample below (above) $0.1 \Zsolar$ is 63\%
(37\%), and the 16th (84th) percentiles of the pLLS metallicities are
$10^{-2.9}$ ($10^{-0.5}$) $\Zsolar$.  Our median metallicities are
lower than \citet{haf17} (${\rm Z} = 10^{-0.9} \Zsolar$ in their
Fig. 9), although we do find a larger dispersion of metallicities than
the FIRE simulations.  Like us, \citet{haf17} finds a shallow
metallicity dependence between $10^{16.2}-10^{19.0} \cms$.

The average offset of our metallicities from \citet{wot16},
$\Delta {\rm Z} \equiv {\rm Z}_{\rm SMOHALOS}- {\rm Z}_{\rm Wotta}$,
is $-0.07$ logarithmic dex for pLLSs and $-0.31$ dex for LLSs
indicating great agreement for the former and a slight under-estimate
in simulations for the latter.  We also quote the typical rms
dispersion in $\Delta {\rm Z}$ between these two samples at $0.21$ and
$0.40$ dex.  Like \citet{wot16}, we perform statistical tests relating
to bimodality on the SMOHALOS pLLS realizations.  They found the
Gaussian Mixture Model \citep[GMM;][]{mur10} rejected a unimodal
distribution at a $>99.9\%$ confidence level and the dip statistic
\citep{har85} predicts a $95.3\%$ chance that the observed
distribution is bimodal or multimodal.  We find the GMM rejects a
unimodal distribution at a $>99.9\%$ confidence level 41\% of the
time.  Interestingly, the highest dip statistic in 100 SMOHALOS
realization is 91.2\% and the median is 20.2\%.  In fact, upon running
1000 realizations, we find only one as high as \citet{wot16}.
Therefore, we conclude that while our realizations often reject
unimodal distributions according to the GMM, they almost never
reproduce the same degree of multimodality from the dip statistic.

We argue that a larger sample size for pLLSs (44) is necessary before
a bimodal metal distribution can be statistically confirmed.  For
example, a CMDF with a uniform distribution between
${\rm Z}= 10^{-2.5}-1 \Zsolar$ would also fall within the 95\%
confidence limits of the observations.  However, we also note that the
degree of bimodality in \citet{wot16} is not exactly reproduced by any
of our realizations, which makes expanding the observed sample important.  

Figure \ref{fig:Wotta-SMOHALOS} also shows SMOHALOS realizations using
a fixed redshift of $z=1$ and $z=0$ but the same distribution of
column densities to demonstrate that redshift evolution is a
significant factor in our simulated distribution.  We predict a
$5\times$ increase in median pLLS metallicity from $10^{-1.6} \Zsolar$
at $z=1$ to $10^{-0.9} \Zsolar$ at $z=0$.  We find a smaller evolution
(0.2 dex) when we simulate the \citet{wot16} pLLS CMDFs in two equally
sized redshift bins split at $z=0.519$; however the \citet{wot16} data
has {\it higher} metallicities in the high-$z$ sample with 10 of 14
pLLSs at $Z\geq 10^{-0.5} \Zsolar$ arising at $z>0.519$.  Our
simulated metallicity evolution that differs from observations is a
central reason why we fail to simulate pLLS bimodality, and we argue
sampling the metallicity evolution of shielded and partially shielded
$\HI$ absorbers over the last 8 Gyrs of cosmic history should continue
to be an observational priority.

\section{Summary and conclusions} \label{sec:dend}

We have explored the metallicities of $\HI$ gas in the EAGLE HiRes
volume across cosmic time from $z=5\rightarrow 0$.  Our results
indicate a steady increase in the metallicities of LLSs and DLAs over
the nearly 13 Gyrs of cosmic time spanned by our study reflecting the
growing stellar content of the Universe.  DLAs are more metal rich
than LLSs in general, although there exists a relatively flat
dependence of metallicity across the range of column densities
spanning from pLLSs to SLLSs.  We predict that on the order of 5\% of
metals nucleosynthesized in and released from stars end up in
$\HI$-traced gas at $z=0$.

Comparisons with observations show some encouraging agreement when
considering the $z\approx 3$ metallicity of LLSs \citep{fum16}, as
well as the evolution of the median metallicity of DLAs \citep{raf14}
and LLSs \citep{leh16,wot16}.  We predict a significant population of
low-metallicity, ${\rm Z}<10^{-3} \Zsolar$ DLAs that are not observed,
which could indicate widespread, early enrichment that is not properly
simulated in EAGLE.  More observations of high-$z$ DLAs could provide
crucial constraints on the enrichment by primordial galaxies and
stars.  We show adequate matches with the median metallicity and the
spread of the $z=1\rightarrow 0$ pLLS distribution, although we do not
reproduce the bimodal metallicity distribution observed by
\citet{wot16}.  We predict more observations will yield more
$\approx 0.1 \Zsolar$ pLLSs and higher pLLS metallicities as evolution
proceeds from $z=1\rightarrow 0$.  SLLS metallicity appears to be
under-predicted by our simulations, while $z<1.5$ DLA metallicity is
over-predicted.

This work represents a first step in using the EAGLE model to
understand the enriched content of the $\HI$-traced Universe.  We will
be working to link these metals to their source galaxies using our
self-consistent, evolutionary model contained within EAGLE.  We create
a URL, http://www.colorado.edu/casa/h1metals, for users to access the
simulation data and data from these plots.

%

\section*{Acknowledgments}
\addcontentsline{toc}{section}{Acknowledgments}

We thank the EAGLE consortium for making the simulations available.
The authors acknowledge Zachary Hafen, Nicolas Lehner, and Joop Schaye
for reading this manuscript and offering helpful suggestions.  We also
wish to thank Chris Wotta for providing us the software and expertise
to run statistical tests on our simulated datasets.  BDO's contribution to
this work was supported by the Hubble Theory grant HST-AR-13262.


%

%


\begin{thebibliography}{}  
%
\bibitem[\protect\citeauthoryear{Adelberger et al.} {2003}]{ade03}Adelberger, K. L., Steidel, C. C., Shapley, A. E., \& Pettini, M. 2003, ApJ, 584, 45
\bibitem[Asplund et al.(2009)]{asp09} Asplund, M., Grevesse, N., Sauval, A. J., Scott, P. 2009, ARA\&A, 47, 481
\bibitem[Bird et al.(2014)]{bir14} Bird, S., Vogelsberger, M., Haehnelt, M., et al.\ 2014, \mnras, 445, 2313
\bibitem[Blitz \& Rosolowsky(2006)]{bli06} Blitz, L., \& Rosolowsky, E.\ 2006, \apj, 650, 933 
\bibitem[\protect\citeauthoryear{Bouch\'e et al.} {2007}]{bou07} Bouch\'e, N., Lehnert, M. D., Aguirre, A., et al.\ 2007, MNRAS, 378, 525
  \bibitem[Cen(2012)]{cen12} Cen, R.\ 2012, \apj, 748, 121 
\bibitem[Chen \& Mulchaey(2009)]{che09} Chen, H.-W., \& Mulchaey, J.~S.\ 2009, \apj, 701, 1219 
\bibitem[\protect\citeauthoryear{Crain et al.}{2015}]{cra15} Crain, R.~A., Schaye, J., Bower, R.~G., et al.\ 2015, MNRAS, 450, 1937
  \bibitem[Crain et al.(2017)]{cra17} Crain, R.~A., Bah{\'e}, Y.~M., Lagos, C.~d.~P., et al.\ 2017, \mnras, 464, 4204 
\bibitem[\protect\citeauthoryear{Dalla Vecchia \& Schaye}{2012}]{dal12} Dalla Vecchia, C., \& Schaye, J.\ 2012, MNRAS, 426, 140
\bibitem[Danforth et al.(2016)]{dan16} Danforth, C.~W., Keeney, B.~A., Tilton, E.~M., et al.\ 2016, \apj, 817, 111 
\bibitem[Dav{\'e} \& Oppenheimer(2007)]{dav07} Dav{\'e}, R., \& Oppenheimer, B.~D.\ 2007, \mnras, 374, 427 
\bibitem[Faucher-Gigu{\`e}re \& Kere{\v s}(2011)]{fau11} Faucher-Gigu{\`e}re, C.-A., \& Kere{\v s}, D.\ 2011, \mnras, 412, L118 
\bibitem[\protect\citeauthoryear{Fukugita \& Peebles} {2004}]{fuk04} Fukigita, M. \& Peebles, P. J. E. 2004, ApJ, 616, 643
\bibitem[Fumagalli et al.(2011)]{fum11} Fumagalli, M., Prochaska, J.~X., Kasen, D., et al.\ 2011, \mnras, 418, 1796 
\bibitem[Fumagalli et al.(2016)]{fum16} 
	Fumagalli, M., O'Meara, J.~M., \& Prochaska, J.~X.\ 2016, \mnras, 455, 4100 
\bibitem[\protect\citeauthoryear{Furlong et al.}{2015}]{fur15} Furlong, M., Bower, R.~G., Theuns, T., et al.\ 2015, MNRAS, 450, 4486 
        %
  \bibitem[\protect\citeauthoryear{Haardt \& Madau}{2001}]{HM01} 
	Haardt F., Madau P.,\ 2001, in Clusters of Galaxies and the High Redshift Universe Observed in X-rays, Neumann D. M., Tran J. T. V., eds.
\bibitem[\protect\citeauthoryear{Hartigan \& Hartigan}{1985}]{har85} Hartigan, J. A., \& Hartigan, P. M. 1985, AnSta, 13, 70
\bibitem[Hafen et al.(2017)]{haf17} Hafen, Z., Faucher-Gigu{\`e}re, C.-A., Angl{\'e}s-Alc{\'a}zar, D., et al.\ 2017, \mnras, 469, 2292 
        %
\bibitem[Katz et al.(1996)]{kat96} Katz, N., Weinberg, D.~H., Hernquist, L., \& Miralda-Escude, J.\ 1996, \apjl, 457, L57 
\bibitem[Lehner et al.(2007)]{leh07} Lehner, N., Savage, B.~D., Richter, P., et al.\ 2007, \apj, 658, 680 
      \bibitem[Lehner et al.(2013)]{leh13} Lehner, N., Howk, J.~C., Tripp, T.~M., et al.\ 2013, \apj, 770, 138 
      \bibitem[Lehner et al.(2016)]{leh16} Lehner, N., O'Meara, J.~M., Howk, J.~C., Prochaska, J.~X., \& Fumagalli, M.\ 2016, \apj, 833, 283
      \bibitem[McAlpine et al.(2016)]{mca16} McAlpine, S., Helly, J.~C., Schaller, M., et al.\ 2016, Astronomy and Computing, 15, 72
\bibitem[Muratov \& Gnedin(2010)]{mur10} Muratov, A.~L., \& Gnedin, O.~Y.\ 2010, \apj, 718, 1266 
\bibitem[Ocvirk et al.(2008)]{ocv08} Ocvirk, P., Pichon, C., \& Teyssier, R.\ 2008, \mnras, 390, 1326 
\bibitem[\protect\citeauthoryear{Oppenheimer et al.}{2016}]{opp16} Oppenheimer, B.~D., Crain, R.~A., Schaye, J., et al.\ 2016, MNRAS, 460, 2157
  \bibitem[\protect\citeauthoryear{Oppenheimer et al.}{2018}]{opp18} Oppenheimer, B.~D., Schaye, J., Crain, R., et al.\ 2017, submitted
\bibitem[\protect\citeauthoryear{Pawlik \& Schaye}{2008}]{Pawlik08} 
  	Pawlik, A.~H., \& Schaye, J.\ 2008, \mnras, 389, 651
%
      \bibitem[\protect\citeauthoryear{Pawlik \& Schaye}{2011}]{Pawlik11} 
  	Pawlik, A.~H., \& Schaye, J.\ 2011, \mnras, 412, 1943 
%
      \bibitem[\protect\citeauthoryear{Peeples et al.}{2014}]{pee14} Peeples, M.~S., Werk, J.~K., Tumlinson, J., et al.\ 2014, ApJ, 786, 54 
\bibitem[\protect\citeauthoryear{Planck Collaboration}{2014}]{pla14} Planck Collaboration, 2014, A\&A, 571, A1
\bibitem[Pontzen et al.(2008)]{pon08} Pontzen, A., Governato, F., Pettini, M., et al.\ 2008, \mnras, 390, 1349 
\bibitem[\protect\citeauthoryear{Prochaska et al.}{2017}]{pro17} Prochaska, J.~X., Werk, J.~K., Worseck, G., et al.\ 2017, ApJ, 837, 169 

      \bibitem[Rafelski et al.(2014)]{raf14}
	Rafelski, M., Neeleman, M., Fumagalli, M., Wolfe, A.~M., \& Prochaska, J.~X.\ 2014, \apjl, 782, L29 
%
  \bibitem[Rahmati et al.(2013a)]{Rahmati13a} 
  	Rahmati, A., Pawlik, A.~H., Rai\v{c}evi\`{c}, M., \& Schaye, J.\ 2013a, \mnras, 430, 2427 
  \bibitem[Rahmati et al.(2013b)]{Rahmati13b} 
  	Rahmati, A., Schaye, J., Pawlik, A.~H., \& Rai\v{c}evi\`{c}, M.\ 2013b, \mnras, 431, 2261 
  \bibitem[Rahmati \& Schaye(2014)]{Rahmati14} 
  	Rahmati, A., \& Schaye, J.\ 2014, \mnras, 438, 529
  \bibitem[Rahmati et al.(2015)]{Rahmati15}
  	Rahmati, A., Schaye, J., Bower, R.~G., et al.\ 2015, \mnras, 452, 2034
  \bibitem[Rahmati et al.(2016)]{Rahmati16}
  	Rahmati, A., Schaye, J, Crain, R.~A., et al.\ 2016, \mnras, 459, 310
  \bibitem[Rai\v{c}evi\`{c} et al.(2013)]{Raicevic13} 
  	Rai\v{c}evi\`{c}, M., Pawlik, A.~H., Schaye, J. \& Rahmati, A.\ 2014, \mnras, 437, 2816
\bibitem[Rosas-Guevara et al.(2015)]{ros15} Rosas-Guevara, Y.~M., Bower, R.~G., Schaye, J., et al.\ 2015, MNRAS, 454, 1038  
\bibitem[\protect\citeauthoryear{Schaller et al.} {2015}]{schal15} Schaller, M., Dalla Vecchia, C., Schaye, J., et al.\ 2015, MNRAS, 454, 2277 
\bibitem[\protect\citeauthoryear{Schaye et al.} {2003}]{sch03} Schaye, J., Aguirre, A., Kim, T.-S., Theuns, T., Rauch, M., \& Sargent, W.L.W. 2003, ApJ, 596, 768
\bibitem[\protect\citeauthoryear{Schaye}{2004}]{sch04} Schaye, J.\ 2004, ApJ, 609, 667 
  \bibitem[\protect\citeauthoryear{Schaye \& Dalla Vecchia}{2008}]{sch08} Schaye, J., \& Dalla Vecchia, C.\ 2008, MNRAS, 383, 1210 
  \bibitem[\protect\citeauthoryear{Schaye et al.}{2015}]{sch15} Schaye, J., Crain, R.~A., Bower, R.~G., et al.\ 2015, MNRAS, 446, 521
  \bibitem[\protect\citeauthoryear{Segers et al.}{2016}]{seg16a} Segers, M.~C., Crain, R.~A., Schaye, J., et al.\ 2016, MNRAS, 456, 1235
\bibitem[\protect\citeauthoryear{Springel} {2005}]{spr05} Springel, V. 2005, MNRAS, 364, 1105
  \bibitem[Tescari et al.(2009)]{tes09} Tescari, E., Viel, M., Tornatore, L., \& Borgani, S.\ 2009, \mnras, 397, 411 
      \bibitem[\protect\citeauthoryear{Trayford et al.}{2015}]{tra15} Trayford, J.~W., Theuns, T., Bower, R.~G., et al.\ 2015, MNRAS, 452, 2879 
\bibitem[Turner et al.(2016)]{tur16} Turner, M.~L., Schaye, J., Crain, R.~A., Theuns, T., \& Wendt, M.\ 2016, MNRAS, 462, 2440
  \bibitem[Turner et al.(2017)]{tur17} Turner, M.~L., Schaye, J., Crain, R.~A., et al.\ 2017, MNRAS, 471, 690 
\bibitem[van de Voort \& Schaye(2012)]{van12} van de Voort, F., \& Schaye, J.\ 2012, \mnras, 423, 2991 
\bibitem[van de Voort et al.(2012)]{van12b} van de Voort, F., Schaye, J., Altay, G., \& Theuns, T.\ 2012, \mnras, 421, 2809 
\bibitem[Weymann et al.(1998)]{wey98} Weymann, R.~J., Jannuzi, B.~T., Lu, L., et al.\ 1998, \apj, 506, 1 
\bibitem[\protect\citeauthoryear{Wiersma et al.} {2009a}]{wie09a} Wiersma, R.~P.~C., Schaye, J., \& Smith, B.~D.\ 2009, MNRAS, 393, 99 
\bibitem[\protect\citeauthoryear{Wiersma et al.} {2009b}]{wie09b} Wiersma, R.~P.~C., Schaye, J., Theuns, T., Dalla Vecchia, C., \& Tornatore, L.\ 2009, 399, 574
  \bibitem[Wotta et al.(2016)]{wot16} Wotta, C.~B., Lehner, N., Howk, J.~C., O'Meara, J.~M., \& Prochaska, J.~X.\ 2016, \apj, 831, 95 

\end{thebibliography}
\end{document}